\documentclass[twocolumn,aps,preprintnumbers,prl,showpacs,amsfonts,amsmath,amssymb,superscriptaddress,floatfix]{revtex4-1}

\usepackage{graphicx}
\usepackage{dcolumn}
\usepackage{bm}

\newcommand \bs{\begin{subequations}}
\newcommand \es{\end{subequations}}
\newcommand \bea{\begin{eqnarray}}
\newcommand \eea{\end{eqnarray}}
\newcommand \be{\begin{equation}}
\newcommand \ee{\end{equation}}

\begin{document}
\title{Mapping the local density of optical states of a photonic crystal  with single quantum dots}
\author{Qin Wang}
\author{S{\o }ren Stobbe}
\author{Peter Lodahl}
\affiliation{DTU Fotonik, Department of Photonics Engineering, Technical University of
Denmark, {\O }rsteds Plads 343, 2800 Kgs.~Lyngby, Denmark.}

\begin{abstract}
 We use single self-assembled InGaAs quantum dots as internal probes to map the local density of optical states of photonic crystal membranes. The employed technique separates contributions from non-radiative recombination and spin-flip processes by properly accounting for the role of the exciton fine structure. We observe inhibition factors as high as 55 and compare our results to local density of optical states calculations available from the literature, thereby establishing a quantitative understanding of photon emission in photonic crystal membranes.

PACS number(s): 42.50.Ct, 78.67.Hc, 78.47.D-

\end{abstract}

\maketitle

Photonic crystals (PCs) are artificial periodic dielectric materials that were originally proposed as a way to dynamically control spontaneous emission of light potentially leading to efficient light sources and solar cells \cite{yabl}. Embedding light sources inside photonic crystals has enabled the experimental demonstration of modified spontaneous emission using dye molecules \cite{koen0,niko1}, quantum wells \cite{fuji} or quantum dots \cite{peter1,finl2,jeppe,niko2,noda}. The complexity of light-matter interaction in PCs is apparent since the experiments combine inhomogeneous dielectric materials varying on the scale of the wavelength with inherently mesoscopic quantum emitters. Within the validity of the dipole approximation, which is valid for standard-sized quantum dots in dielectric structures \cite{jeppe2}, the light-matter coupling is determined by two quantities: i) the transition dipole moment of the optical transition and ii) the local density of optical states (LDOS) projected onto the orientation of the dipole \cite{spri,busc}. The latter is a property of the PC that accounts for all available Bloch modes at a specific position and has been exceedingly challenging to calculate in realistic structures relevant for experiments \cite{koen1,busc}. Consequently, a thorough understanding of the LDOS and therefore the potential of spontaneous emission control has been lacking in PCs. Here we present the experimental mapping of the LDOS of PCs by time-resolved emission studies of single quantum dots (QDs) with well-characterized optical properties.

The focus of the present work is to use QDs as internal probes of the LDOS. This quest requires a detailed understanding of the optical properties of QDs in inhomogeneous media and in particular the role of the exciton fine structure. Our previous work on dielectric interfaces demonstrated the necessity to consider the role of dark excitons \cite{peter2} and proved the validity of the dipole approximation for standard-sized QDs in dielectric structures \cite{jeppe2}, while a breakdown of the dipole approximation was observed near metallic interfaces \cite{ande}. Building on this knowledge, we present a new and general method to extract the \emph{radiative} decay rate of single QDs as opposed to the \emph{total} decay rate that is otherwise directly measured in time-resolved spectroscopy and is significantly influenced by dark excitons and other non-radiative decay processes. From the radiative decay rate we determine the LDOS and experimentally map it by recording decay curves of many single QDs positioned throughout the 2D PC membranes. We compare our results with LDOS simulations performed with 3D finite-difference time-domain (FDTD) simulations available in the literature \cite{koen1} and explicitly demonstrate the importance of extracting the radiative decay rate. We observe record high spontaneous emission inhibition factors of $55$ compared to QDs in homogenous media, which proves the potential of 2D PC membranes for applications where spontaneous emission is a nuisance. The frequency dependency of the LDOS is mapped out both inside and outside the 2D band gap, and in the latter case also enhancement of the spontaneous emission rate is observed.

Within the dipole approximation and for weak light-matter interaction strengths, the radiative decay rate is directly proportional to the LDOS: $\gamma _{\rm{rad}} (\mathbf{r_0} ,\omega ) = \frac{{\pi \omega }}{{3\hbar \varepsilon _0 }}\left| \boldsymbol{\mu} \right|^2 \rho _\mu  (\mathbf{r_0} ,\omega )$, where $\rho _ \mu (\mathbf{r_0} ,\omega )$ is the projected LDOS evaluated at the position $\mathbf{{r}_0}$ and emission frequency $\omega$ of the emitter, $\varepsilon _0$ is the vacuum permittivity, and $\boldsymbol{\mu} $ is the transition dipole moment. The LDOS describes the electromagnetic environment and can be calculated from the dyadic Green's function \cite{novo}. Successfully extracting the radiative decay rate provides a mean to obtaining the LDOS by comparing to the radiative decay rate in a homogeneous medium $\gamma_\mathrm{rad}^{\hom}(\omega)$ using the relation
\begin{align}
\rho _\mu  ({\bf{r}}_{\bf{0}} ,\omega ) = \frac{{\gamma _{{\rm{rad}}} ({\bf{r}}_{\bf{0}} ,\omega )}}{{\gamma _{{\rm{rad}}}^{\hom }(\omega ) }}\rho (\omega ),
\end{align}
where $\rho (\omega ) = \frac{{n\omega ^2 }}{{3 \pi ^2 c^3 }}$ is the projected LDOS for a homogeneous medium with refractive index $n$ and $c$ is the speed of light in vacuum.

A detailed understanding of the exciton fine structure and the role of non-radiative recombination is required in order to use QDs as LDOS probes \cite{jeppe2,soren2,peter2}. An InGaAs QD has two optically active bright states with total angular momentum $J_z  =  \pm 1$ and two dark states with $J_z  =  \pm 2$, where the quantization axis $z$ is the growth direction [001] \cite{baye}. Due to the reduced symmetry and anisotropic exchange interactions, the two bright states are separated in energy and form two eigenstates X and Y named according to their dipole orientations ($[110]$ or [$1\overset{-}{1}0$]), where $\left \vert
X\right \rangle _\mathrm{b}=(\left \vert +1\right \rangle +\left \vert -1\right \rangle
)/\sqrt{2}$, $\left \vert Y\right \rangle _\mathrm{b}=(\left \vert +1\right \rangle
-\left \vert -1\right \rangle )/\sqrt{2}$. Similarly, the two dark states are separated into $\left \vert X\right \rangle
_\mathrm{d}=(\left \vert +2\right \rangle +\left \vert -2\right \rangle )/\sqrt{2}$,
$\left \vert Y\right \rangle _\mathrm{d}=(\left \vert +2\right \rangle -\left \vert
-2\right \rangle )/\sqrt{2}$. The PC patterns can be aligned relative to X and Y, as schematically indicated in Fig.~1(a).

\begin{figure}[ptb]
\includegraphics[scale=0.535]{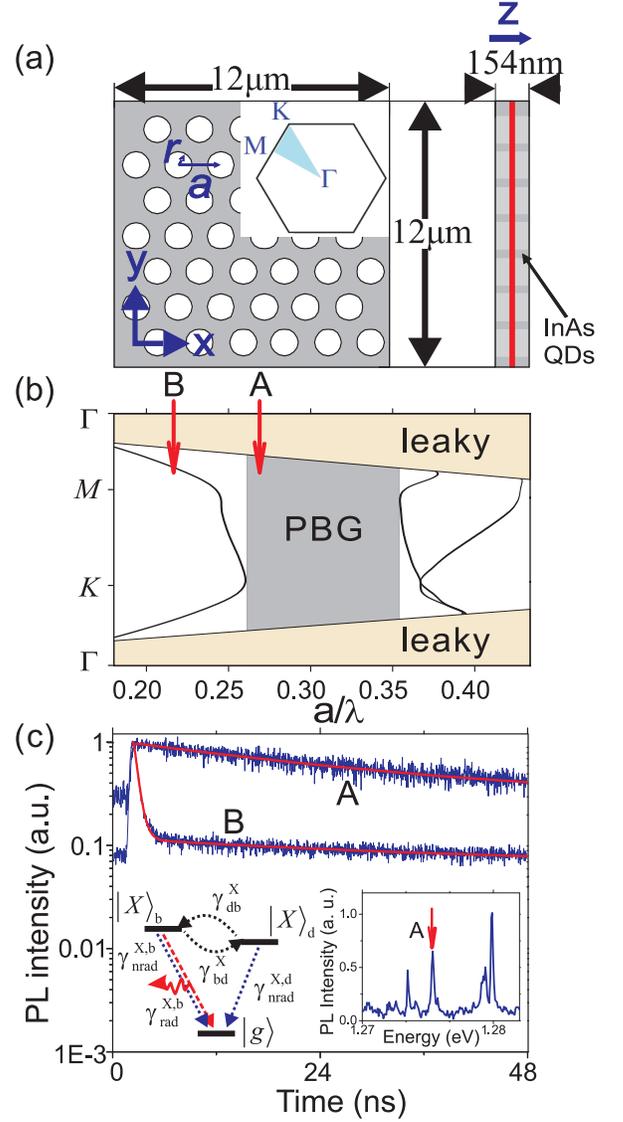}
\caption{Fig.~1 (Color online) (a) Sketch of a PC membrane, top view (left) and side view (right). The orientations of the X and Y dipoles together with the sample growth direction (Z) are schematically indicated. The inset shows the first Brillouin zone with indications of the relevant symmetry directions; (b) A photonic-band diagram calculated for optical modes with in-plane polarization, where the gray region represents the 2D photonic band gap (PBG) and "leaky" refers to the regions where light is not confined to the membrane; (c) Measured decay curves of QD A and B that are, respectively, tuned in- or outside the band gap, as schematically indicated in (b). The insets show the three-level diagram of the QD (left) and a measured emission spectrum (right).}
\label{Fig1}
\end{figure}

The two bright and two dark states together with the ground state form a five-level model. $\left| X \right\rangle _\mathrm{b}$ couples to $\left| X \right\rangle _\mathrm{d}$ through either an electron or a hole spin flip mediated by phonons and exchange interaction \cite{rosz}. We note that spin-flip processes coupling bright excitons, i.e. $\left| X \right\rangle _\mathrm{b} \leftrightarrows \left| Y \right\rangle _\mathrm{b} $ are slow compared to the other decay processes and therefore can be abandoned in the analysis, as theoretically predicted \cite{rosz,tsit} and experimentally confirmed from the large anisotropy in the decay rate for X and Y states observed in a PC \cite{qin1}. Consequently, the five-level scheme can be simplified to the three-level system indicated in Fig.~1(c). The bright state  can decay either through radiative or non-radiative processes with rates $\gamma_\mathrm{rad}^{X,b}$ and $\gamma_\mathrm{nrad}^{X,b}$, respectively, and is coupled to the dark state through a spin-flip rate $\gamma_\mathrm{bd}^{X}$. Radiative transitions from the dark state to the ground state are forbidden, but non-radiative recombination is possible with the rate $\gamma_\mathrm{nrad}^{X,d}$ together with spin flips to the bright state with a rate $\gamma_\mathrm{db}^{X}$. Solving the resulting rate equations we find that the population of bright excitons decay bi-exponentially $\rho_\mathrm{b}^{X}(t)=A_\mathrm{f}e^{-\gamma_\mathrm{f}^Xt}+A_\mathrm{s}e^{-\gamma_\mathrm{s}^Xt}$, with $\gamma _\mathrm {f}^X  = \gamma _\mathrm{nrad}^{X,b}  + \frac{1}{2} \gamma _\mathrm{rad}^{X,b}  + \gamma _\mathrm{db}^X  + \sqrt {(\gamma _\mathrm{db}^X )^2 + (\gamma _\mathrm{rad}^{X,b}/2 )^2}$, $\gamma _\mathrm{s}^X  = \gamma _\mathrm{nrad}^{X,b}  + \frac{1}{2} \gamma _\mathrm{rad}^{X,b}  + \gamma _\mathrm{db}^X  - \sqrt {(\gamma _\mathrm{db}^X )^2 + (\gamma _\mathrm{rad}^{X,b}/2 )^2}$, $A_\mathrm{f}^X  = \frac{\rho _\mathrm{b}^X (0)}{2} \left[1 + \frac{\gamma _\mathrm{rad}^{X,b}}{\gamma _\mathrm {f}^X-\gamma _\mathrm {s}^X} \right] - \rho _\mathrm{d}^X (0)  \frac{\gamma _\mathrm{db}^{X}}{\gamma _\mathrm {f}^X-\gamma _\mathrm {s}^X}$, and $A_\mathrm{s}^X  = \frac{\rho _\mathrm{b}^X (0)}{2} \left[1 - \frac{\gamma _\mathrm{rad}^{X,b}}{\gamma _\mathrm {f}^X-\gamma _\mathrm {s}^X} \right] + \rho _\mathrm{d}^X (0)  \frac{\gamma _\mathrm{db}^{X}}{\gamma _\mathrm {f}^X-\gamma _\mathrm {s}^X}.$ Similar expressions can be obtained for $\rho_\mathrm{b}^{Y}(t).$  $\rho_\mathrm{b}^{X}(0)$ and $\rho_\mathrm{d}^{X}(0)$ are the initial populations of the bright and dark states. For non-resonant and weak pumping we have $\rho_\mathrm{d}^{X}(0)=\rho_\mathrm{b}^{X}(0)=0.5,$ where we note that the presented results of the LDOS mapping are robust to deviations from  this equilibrium condition. Furthermore, we have assumed: (i) $\gamma_\mathrm{bd}^{X}=\gamma_\mathrm{db}^{X}$ and (ii)
$\gamma_\mathrm{nrad}^{X,b}=\gamma_\mathrm{nrad}^{X,d}$. (i) is a good approximation for
intermediate temperatures ($T=10$ K in our experiment), because
$\gamma_\mathrm{bd}=e^{\delta_\mathrm{bd}/k_\mathrm{B}T}\gamma_\mathrm{db}$, and $k_\mathrm{B}T\gg \delta_\mathrm{bd}$,
where $k_\mathrm{B}$ is the Boltzmann constant, and $\delta_\mathrm{bd}$ is the
energy splitting between the bright and dark states (typically, a few hundred $\mu$eV \cite{baye}). Assumption (ii) has been proven valid by experiments on QDs in dielectric media with a known LDOS \cite{soren2,peter2}.
The intensity measured in a time-resolved experiment
is $I^{X}(t)=C_\mathrm{0}\gamma_\mathrm{rad}^{X,b}\rho_\mathrm{b}^{X}(t)$,
where $C_\mathrm{0}$ is proportional to the total collection efficiency of the experimental setup, which is an overall scaling factor of the decay curves. After fitting each decay curve with a bi-exponential function, we obtain the four parameters $\gamma_\mathrm{f}^i$, $\gamma_\mathrm{s}^i$, $\ A_\mathrm{f}^i$,
and $\ A_\mathrm{s}^i$ (for $i=X,Y)$ and  $\gamma_\mathrm{rad}^{i}$,
$\gamma_\mathrm{nrad}^{i}$ and $\gamma_\mathrm{bd}^{i}$ can be extracted from the relations described above. The amplitudes need to be corrected to account for residual population of the QD when it is re-excited by a light pulse \cite{peter2}:  $A_\mathrm{f}^i  = \tilde A_\mathrm{f}^i [1 - \exp ( - \gamma _\mathrm{f}^i \tau)]$ and $A_\mathrm{s}^i  = \tilde A_\mathrm{s}^i [1 - \exp ( - \gamma _\mathrm{s}^i \tau)]$, where $\tau$ is the excitation period, and $\tilde A_\mathrm{f}^i$ ($\tilde A_\mathrm{s}^i$) refers to the measured amplitude. We stress that the presented method is completely general, i.e., no assumptions about the magnitude of the rates have been implemented, and therefore InGaAs QDs can be employed as LDOS probes in any nanophotonic environment.

We have carried out time-resolved measurements on single self-assembled QDs positioned in- or outside 2D PC membranes.
The experiments are done in a flow cryostat at $10$ K and the sample consists of a series of 2D GaAs (n=3.5) PC membranes with a layer of InGaAs QDs (density $\sim$ 80 $\mu$m$^{-2}$) embedded in the center. The dimension of each PC membrane is $12 \times 12$ $\mu$m$^2$ with a thickness of 154 nm as shown in Fig.~1(a) where also the orientations of X and Y dipoles are indicated. Fig.~1(b) shows the photonic band diagram of the structure with indications of the 2D photonic band gap region and the continuum due to coupling to leaky modes out of the membrane.  The lattice parameter ($a$) ranges from 200 to 385 nm in steps of 5 nm, and the $r/a$ ratio is fixed at 0.30, where $r$ is the radius of the air hole. The QDs are excited with a PicoQuant PDL-800 pulsed diode laser at 781 nm with varying repetition rate (5, 10, 20, 40 MHz). Under a weak excitation condition, we identify neutral excitons by their excitation power and polarization dependence, i.e., multi-exciton complexes are excluded due to the observed linear power dependence before saturation, while single-charge excitons are found to be mono-exponential and have a very weak polarization dependence. We select only QDs that emit within a narrow spectral range of $970 \pm 5$ nm, in order to probe QDs with similar oscillator strengths \cite{jeppe2}. In total 88 QDs in the PC are probed in addition to 5 QDs positioned outside the PC pattern that serve as reference. For each QD, a polarizer is used to record emission from either the X or the Y exciton state.

Figure~1(c) shows two typical decay curves of  QD A and B (X states) that are tuned respectively in- or outside the band gap. Clearly, the decay of QD A is strongly suppressed as a consequence of the strongly suppressed LDOS associated with the 2D band gap of the structure. QD B is positioned in a PC that is designed to probe the LDOS at the edge of the band gap, and indeed a much faster decay is observed in this case. Based on the method explained above, we can extract the radiative decay rate of each exciton state and obtain the projected LDOS. The results are shown in Fig.~2 including a comparison to  the theory of Koenderink \textit{et~al.} \cite{koen1} who calculated the projected LDOS for 4 specific positions in a PC membrane. In order to compensate for a slight difference in membrane thickness between experiment and theory, we have extended the band gap width according to the theory of Ref. \cite{andr}. We observe a pronounced 2D band gap leading to a wide range of frequencies where the LDOS is strongly suppressed  for both X and Y exciton states. Large point-to-point fluctuations in the experimental data are observed since the QDs exhibit different radiative decay rates due to their varying positions in the PC membrane, thus reflecting the sensitive spatial variation of the LDOS. Outside the band gap an enhanced LDOS is observed relative to the DOS of a homogeneous medium. Consequently, the decay rate of QDs tuned to the band edge can be Purcell enhanced in this case by coupling to extended Bloch modes as opposed to localized cavity modes, which is the most common way of realizing the Purcell effect. In general good agreement between experiment and theory is observed, especially taking into consideration that the spatial sampling used in the theory is rather sparse, while in the experiment the QDs occupy all different positions in the high refractive index material. Interestingly, the experimental data appear to be biased such that the measured LDOS seems systematically larger than predicted by theory, in particular for normalized frequencies below the 2D band gap. This observation is likely an effect of unavoidable fabrication disorder that would be more severe for the small feature sizes corresponding to reduced frequencies below the band gap. Disorder has been found to lead to significant modifications of the LDOS in PC waveguides \cite{luca} and it is therefore important to quantify the role of disorder in any PC application, which is done here through the comparison with theory.

\begin{figure}[ptb]
\includegraphics[scale=0.45]{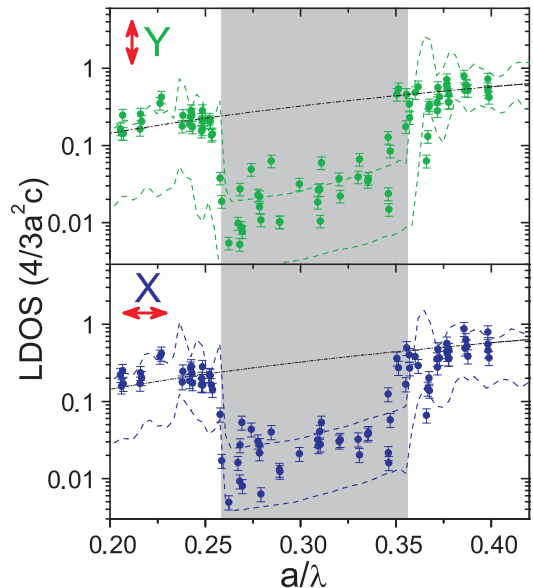}
\caption{Fig. 2 (Color online) Measured projected LDOS of PC membranes versus normalized emission frequencies for X and Y dipole orientations (data points). The dashed lines represent the calculated projected LDOS as taken from Ref. \cite{koen1}. For reference, also the DOS of homogeneous GaAs is indicated (black dash-dotted lines) and the gray areas represent the band gap.}
\label{Fig2}
\end{figure}

\begin{figure}[ptb]
\includegraphics[scale=0.45]{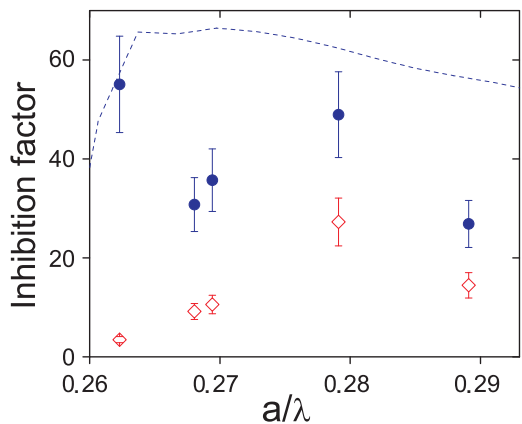}
\caption{Fig. 3 (Color online) Inhibition factors for five strongly inhibited QDs due to the 2D photonic band gap. The blue circular points are obtained from the full model accounting for the exciton fine structure and non-radiative recombination, while the red diamond points are obtained by incorrectly using directly measured total decay rates. The dashed blue line shows the maximum values from simulations \cite{koen1}.}
\label{Fig3}
\end{figure}

For reference, we have also measured the radiative decay rate, non-radiative decay rate and spin-flip rate of 5 QDs positioned outside the PC area and obtain the average values  $\gamma _\mathrm{{rad}}^i=1.1 \pm 0.1$ ns$^{-1}$, $\gamma _\mathrm{{nrad}}^i=0.06 \pm 0.05$ ns$^{-1}$ and $\gamma _\mathrm{{bd}}^i=0.005 \pm 0.002$ ns$^{-1}$ respectively, which agree well with previous results \cite{jeppe2,soren2,peter2}. Focusing on QDs positioned in the band gap we extract the inhibition factor relative to the reference value in the homogenous medium, see Fig.~3. A maximum inhibition factor of $55$ is observed, which is the highest value ever reported in any PC. We stress the necessity of employing the presented method that accounts for the QD fine structure and non-radiative recombination in order to correctly extract the inhibition factors. Thus, Fig. 3 also displays the inhibition factors derived by using the common but incorrect assumption that the directly measured total decay rate is dominated by radiative recombination. This comparison clearly illustrates the importance of employing the correct microscopic model of the quantum emitter in order to use them as LDOS probes.

In conclusion, we have presented a method to probe the LDOS of any nano environment by employing self-assembled InGaAs QDs. By properly accounting for the exciton fine structure, it is possible to extract the radiative decay rate and therefore eliminate effects from non-radiative recombination and spin-flip processes. We presented a detailed frequency map of the LDOS and a detailed comparison to existing theory. Inhibition factors as high as 55 were observed inside the 2D band gap, thus clearly demonstrating the potential of PC membranes for efficient spontaneous emission inhibition. Our work is expected to lay the foundation for further exploitations of photonic crystal membranes for all-solid-state quantum electrodynamics experiments, where the LDOS is the essential quantity that controls not only spontaneous emission, but also, e.g., the Lamb shift \cite{vats} or Casimir forces.

The authors thank P. T. Kristensen and H. T. Nielsen for useful discussions and A. F. Koenderink for generously sharing the data of his LDOS simulations. We gratefully acknowledge financial support from the Villum Foundation, The Danish Council for Independent Research (Natural Sciences and Technology and Production Sciences), and the European Research Council (ERC consolidator grant).

\end{document}